\documentclass{emulateapj}
\usepackage{lscape, longtable, hyperref}

\newcommand{\lapprox }{{\lower0.8ex\hbox{$\buildrel <\over\sim$}}}
\newcommand{\gapprox }{{\lower0.8ex\hbox{$\buildrel >\over\sim$}}}
\def\amin{\ifmmode^{\prime}\else$^{\prime}$\fi}
\def\asec{\ifmmode^{\prime\prime}\else$^{\prime\prime}$\fi}

\slugcomment{DRAFT \today}
\shorttitle{GBT Observations of Low-Mass WDs }
\shortauthors{Ag{\" u}eros et al.}

\begin{document}

\title{A Radio Search For Pulsar Companions To SDSS Low-Mass White Dwarfs}

\author{
Marcel A.\ Ag\"ueros\altaffilmark{1,2},
Fernando Camilo\altaffilmark{1},
Nicole M.\ Silvestri\altaffilmark{3},
S.\ J.\ Kleinman\altaffilmark{4},
Scott F.\ Anderson\altaffilmark{3},
James W.\ Liebert\altaffilmark{5}
}

\altaffiltext{1}{Columbia Astrophysics Laboratory, Columbia University, New York, NY 10027; marcel@astro.columbia.edu} 
\altaffiltext{2}{NSF Astronomy \& Astrophysics Postdoctoral Fellow}
\altaffiltext{3}{Department of Astronomy, University of Washington, Seattle, WA 98195}
\altaffiltext{4}{Gemini Observatory, Northern Operations Center, Hilo, HI 96720}
\altaffiltext{5}{Steward Observatory, University of Arizona, Tucson, AZ 85121}

\begin{abstract}
We have conducted a search for pulsar companions to $15$ low-mass white dwarfs (LMWDs; M $< 0.4$ M$_\odot$) at $820$ MHz with the NRAO Green Bank Telescope (GBT). These LMWDs were spectroscopically identified in the Sloan Digital Sky Survey (SDSS), and do not show the photometric excess or spectroscopic signature associated with a companion in their discovery data. However, LMWDs are believed to evolve in binary systems and to have either a more massive WD or a neutron star as a companion. Indeed, evolutionary models of low-mass X-ray binaries, the precursors of millisecond pulsars (MSPs), produce significant numbers of LMWDs \citep[e.g.,][]{benvenuto05}, suggesting that the SDSS LMWDs may have neutron star companions. No convincing pulsar signal is detected in our data. This is consistent with the findings of \citet{joeri07}, who conducted a GBT search for radio pulsations at $340$ MHz from unseen companions to eight SDSS WDs (five are still considered LMWDs; the three others are now classified as ``ordinary'' WDs). We discuss the constraints our non-detections place on the probability $P_{MSP}$ that the companion to a given LMWD is a radio pulsar in the context of the luminosity and acceleration limits of our search; we find that $P_{MSP} < 10^{+4}_{-2}\%$. 
\end{abstract}


\section{Introduction}
The number of known white dwarfs (WDs) has increased dramatically since the beginning of the Sloan Digital Sky Survey \citep[SDSS;][]{york00}. The most recent SDSS WD catalog, collected from its Data Release 4 \citep[DR4;][]{DR4paper}, yielded $9300$ spectroscopically confirmed WDs, of which $6000$ are new discoveries \citep{eisenstein06}. Model fits to the SDSS spectra find that the majority of these objects are hydrogen (DA) WDs for which the mass distribution peaks at the expected value of $0.6$ M$_\odot$\footnote{The existence of this peak in the field DA mass distribution has been known for close to $30$ years; see \citet{koester79}.}. But among the new SDSS WDs there are also a handful of candidate very low-mass WDs (LMWDs) with M $< 0.4$ M$_\odot$. Such objects are noteworthy not only because of their rarity but because the Galaxy is not thought to be old enough to produce LMWDs through single star evolution. 

The youngest WDs in the oldest Milky Way globular clusters have masses of $\sim0.5$~M$_\odot$ \citep{hansen07}. Lower mass WDs must therefore undergo significant mass loss, presumably because they form in close binaries whose evolution includes a phase of mass transfer. During this phase, much of the WD progenitor's envelope is removed, stunting the WD's evolution by preventing a helium flash in its core and resulting in the observed low-mass, helium-core WD\footnote{\citet{kilic07c} argue that a LMWD may form from the evolution of a single, metal-rich star. However, they predict that the binary fraction for WDs with M $\sim 0.4$~M$_\odot$ is $50\%$ and rises to $100\%$ for M $< 0.2$~M$_\odot$. Thus, even in this scenario, the likelihood of finding companions to the SDSS LMWDs remains high.}. As \citet{marsh95} put it, ``low-mass WDs need friends.'' In examining seven WDs with M~$\leq 0.45$~M$_\odot$, they found that four are in double-degenerate close binary systems, with orbital periods on the order of a few hours to a few days; for a fifth they identified a companion but were unable to confirm its nature or obtain an orbital period. 
 
Subsequent studies of large samples of DAs found that roughly $10\%$ have masses $\lapprox\ 0.4$~M$_{\odot}$, and that in the majority of these binaries the companion is likely to be a degenerate star \citep{liebert04,liebert05}. The \citet{marsh95} systems are WD/WD binaries, but the LMWD companions can also be neutron stars \citep[NSs; cf.\ discussion in][]{driebe98}. In these systems, binary evolution produces a LMWD in a near-circular orbit around the NS ``reborn'' as a millisecond pulsar \citep[MSP; for discussion of different evolutionary scenarios involving NSs, see][]{tauris06}.

Observationally, roughly $5\%$ of field radio pulsars reside in binary systems. Most of these pulsars are MSPs, and in most cases their companions are thought to be LMWDs with $0.1\ \lapprox$~M $\lapprox\ 0.4$ M$_\odot$. Of the $52$ MSP companions identified outside of globular clusters, \citet{manchester05} list $17$ with masses $\lapprox\ 0.2$~M$_\odot$, assuming the systems have a median inclination of $60^{\circ}$. However, the MSP companions are frequently too faint for optical spectroscopy to confirm that they are LMWDs. Despite the observational challenges, there are several well known MSP/LMWD systems, and in fact the majority of known He-core WDs are MSP companions \citep{panei07}; see \citet{vankerkwijk05} for a review of these systems.

That said, the birthrate of MSPs in the Galactic disk is thought to be $\sim3\times10^{-6}$\,yr$^{-1}$ \citep{lorimer08}. Estimating the number of LMWDs in the Galaxy is difficult \citep[cf.\ discussion in][]{liebert05}, and the SDSS sample is best considered as providing a lower limit (because WDs are generally not explicitly targeted for SDSS spectroscopy, it is unclear how best to correct for e.g., incompleteness). However, \citet{liebert05} used the Palomar Green Survey to estimate the formation rate of LMWDs to be $0.4\times10^{-13}$ pc$^{-3}$ yr$^{-1}$, implying that these make up about $10\%$ of all DA white dwarfs\footnote{This contribution is $\sim10\times$ larger than indicated by the relative space densities of low-mass and ordinary ($0.6$ M$_{\odot}$) WDs as it accounts for the presence of degenerate companions to the LMWDs.}. Naively, this suggests that the LMWD birthrate in the volume sampled by the SDSS DR4 ($\sim4$ kpc$^3$) is $\sim2 \times 10^{-4}$ yr$^{-1}$, or nearly two orders of magnitude higher than that of MSPs. This discrepancy in the (admittedly very rough) birthrates suggests that there is a low likelihood of detecting MSP companions to SDSS LMWDs. 

Still, because of the tantalizing theoretical and observational connections between LMWDs and MSPs, we searched for radio pulsations from putative pulsar companions to $15$ spectroscopically confirmed SDSS LMWDs using the NRAO Green Bank Telescope (GBT). We describe the sample of SDSS LMWDs in \S~\ref{wds} and our $820$ MHz observations in \S~\ref{obs}. We discuss the significance of our non-detections in \S~\ref{discussion} and conclude in \S~\ref{concl}.

\section{The Targets: SDSS Low-Mass WDs}\label{wds}
Our sample of candidate LMWDs was selected from the two SDSS WD catalogs \citep[for a full description of the SDSS WD selection, WD models, and the {\tt autofit} program used to obtain estimates for WD temperatures and masses, see][]{kleinman04,eisenstein06}. Two, SDSS J105611.03$+$653631.5 and J123410.37$-$022802.9, are described in detail in \citet{liebert04}. An additional $11$ candidate LMWDs were later proposed by \citet{eisenstein06}. These WDs are selected to have temperatures T$_{\rm eff} < 30,000$ K, surface gravities (log~$g$) more than $2\sigma$ below $7.2$, and $g < 20$ mag. None of these WDs show the photometric excess or spectroscopic signature associated with a companion in their discovery data \citep[see, for example, Figures 21 and 22 in][]{eisenstein06}, as expected if the companion is a compact object.

\begin{figure*}[th]
\centerline{\includegraphics[width=1.5\columnwidth]{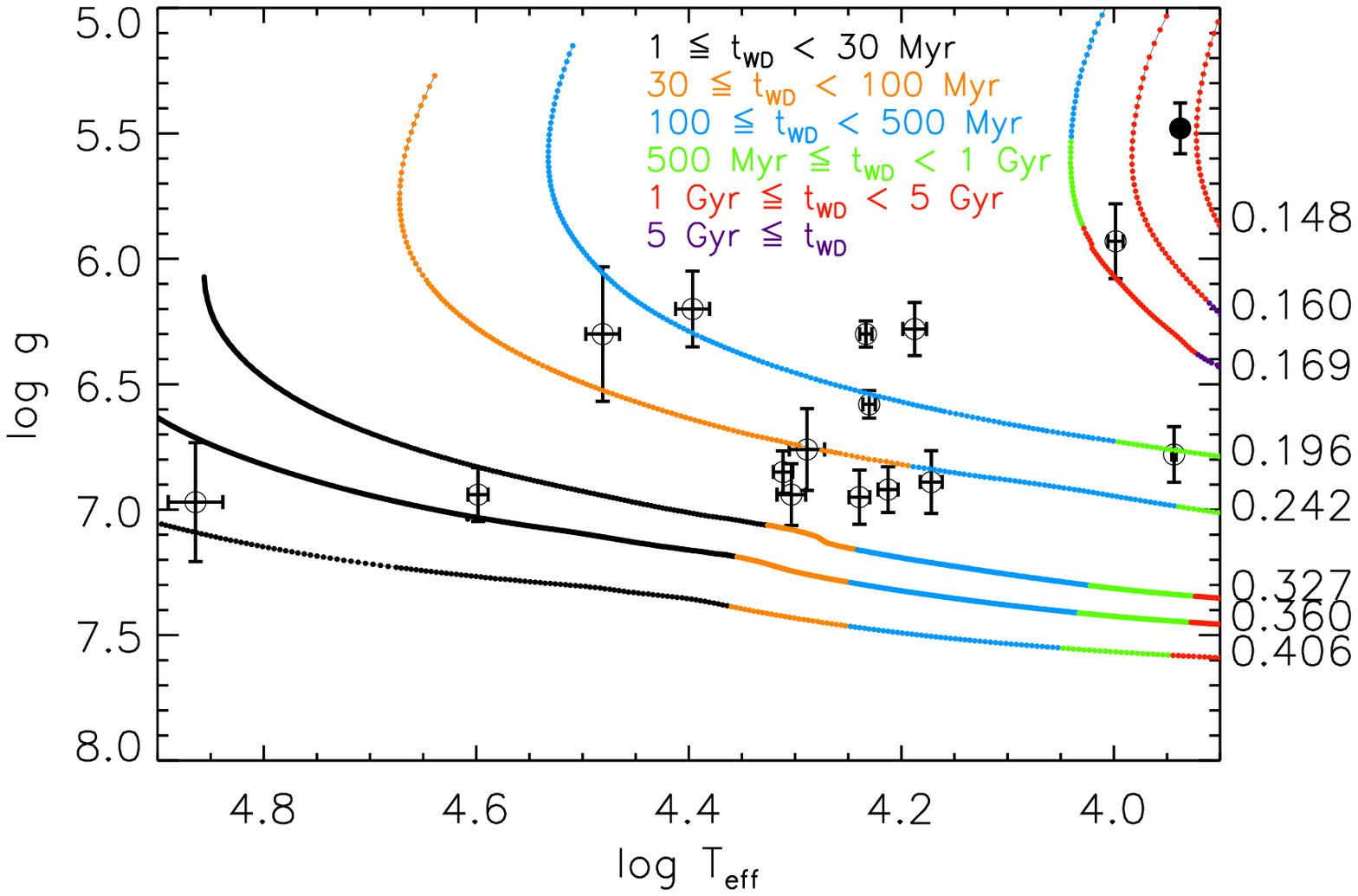}}
\caption{Log~$g$ vs.\ log T$_{\rm eff}$ for the SDSS LMWDs overlaid on tracks of constant mass (in units of M$_\odot$) from \citet{serenelli01} for M $= 0.148$ and $0.160$ M$_\odot$ and from \citet{althaus01} for M~$\ge0.169$ M$_\odot$. A canonical $0.6$ M$_\odot$ DA WD has log~$g$ $= 8$ and the observed peak in the temperature distribution for SDSS DAs is log T$_{\rm eff} \approx 4$. The tracks are color-coded to indicate the age of a LMWD of a given mass, calculated by these authors from the end of mass transfer onto the WD companion. The solid data point is J2049, identified as a A2 star by \citet{kilic07a}. The three WDs with log T$_{\rm eff} > 4.4$ were not included in the \citet{eisenstein06} LMWD list.}\label{properties}
\end{figure*}

Re-analysis of the SDSS spectra by \citet{kilic07a} confirmed that the \citet{eisenstein06} candidates are LMWDs in all but one case: they find that SDSS J204949.78$+$000547.3 is an A2 star\footnote{This work was done after we had observed J2049 and we therefore analyzed these data as well.}. This leaves us with a sample of $12$ LMWDs from the SDSS catalogs. To these we add three LMWD candidates identified by \citet{kleinman04} that are present in the \citet{eisenstein06} WD catalog but are not classified by the latter authors as LMWDs. SDSS J131033.25$+$644032.8 and 234536.47$-$010204.9 are not included in the \citet{eisenstein06} LMWD list because they have T$_{\rm eff} = 39,700$ and $30,300$~K, respectively. The model fits for SDSS J163800.36$+$004717.7 return T$_{\rm eff} = 73,000$~K and a log~$g$ that is only $1\sigma$ below $7.2$. Model fits to very cool and very hot WDs are somewhat unreliable\footnote{For example, differences between SDSS-derived temperatures and those in the literature for previously cataloged WDs are about $10\%$ at $50,000$~K \citep{eisenstein06}.}, but we wish to be as inclusive as possible in assembling our sample. Adding these three gives us just $15$ LMWD candidates spectroscopically identified in a survey volume of $\sim4$~kpc$^3$.

\begin{deluxetable*}{lcccccccc}
\tablewidth{0pt}
\tabletypesize{\scriptsize}
\tablecaption{SDSS LMWDs Observed With The GBT\label{props}}
\tablehead{
\colhead{}       & \colhead{SDSS $g$} & \colhead{T$_{\rm eff}$} & \colhead{M$_{\rm WD}$} & \colhead{$P_{\rm orb}$} & \colhead{Dist.} & \colhead{$b$}        & \colhead{DM}  &\colhead{Integration}  \\
\colhead{SDSS J} & \colhead{(mag)}     & \colhead{(K)}           & \colhead{(M$_\odot$)}  & \colhead{(d)}           & \colhead{(kpc)} & \colhead{($^{\circ}$)} & \colhead{(cm$^{-3}$ pc)} & \colhead{Time (s)}    
}
\startdata
084910.13$+$044528.7 & $19.31\pm0.02$ & $9962\pm165$  & $\sim0.169$ & $\lapprox\ 0.1$ & $1.5$ & $+28.3$ & $112$ & $3000$ \\
123410.37$-$022802.9 & $17.87\pm0.02$ & $17114\pm227$ & $0.18-0.19$ & $\lapprox\ 0.3$ & $0.7$ & $+60.1$ & $64$ & $13400$ \\
105353.89$+$520031.0 & $18.93\pm0.02$ & $15399\pm400$ & $<0.196$    & $\lapprox\ 0.3$ & $1.2$ & $+56.8$ & $64$  & $2400$ \\
082212.57$+$275307.4 & $18.33\pm0.01$ & $8777\pm40$   & $\sim0.196$ & $\sim0.3$       & $0.9$ & $+31.1$ & $128$ & $1800$ \\
143633.29$+$501026.8 & $18.23\pm0.01$ & $16993\pm229$ & $\sim0.196$ & $\sim0.3$       & $0.9$ & $+59.5$ & $64$  & $1500$ \& $1800$ \\
162542.10$+$363219.1 & $19.36\pm0.02$ & $24913\pm936$ & $\sim0.196$ & $\sim0.3$       & $1.5$ & $+44.1$ & $64$  & $3900$ \\
234536.47$-$010204.9\tablenotemark{a} & $19.50\pm0.02$ & $30270\pm1127$ & $<0.242$ & $\sim7.5$ & $1.6$ & $-59.5$ & $64$ & $2400$ \\
002207.65$-$101423.5 & $19.77\pm0.02$ & $19444\pm758$ & $<0.292$    & $\lapprox\ 30$  & $1.8$ & $-71.8$ & $64$ & $2400$ \\
002228.45$+$003115.5 & $19.23\pm0.05$ & $17355\pm394$ & $<0.292$    & $\lapprox\ 30$  & $1.4$ & $-61.5$ & $64$ & $2200$ \\
142601.47$+$010000.2 & $19.34\pm0.02$ & $16311\pm359$ & $<0.292$    & $\lapprox\ 30$  & $1.5$ & $+55.4$ & $64$ & $2400$ \\
163030.58$+$423305.7 & $19.02\pm0.01$ & $14854\pm359$ & $<0.292$    & $\lapprox\ 30$  & $1.3$ & $+43.4$ & $80$ & $1800$ \\
225242.25$-$005626.6 & $18.59\pm0.03$ & $20479\pm433$ & $<0.292$    & $\lapprox\ 30$  & $1.0$ & $-50.0$ & $80$ & $2400$ \\
105611.03$+$653631.5 & $19.77\pm0.03$ & $20112\pm634$ & $\sim0.31$  & $\sim35$        & $1.8$ & $+47.5$ & $80$ & $10400$ \\
131033.25$+$644032.8\tablenotemark{a} & $18.30\pm0.02$ & $39654\pm874$ & $\lapprox\ 0.360$ & $\sim200$ & $0.9$ & $+52.3$ & $64$ & $1300$\\
163800.36$+$004717.7\tablenotemark{a} & $18.84\pm0.01$ & $73149\pm4459$ & $\lapprox\ 0.406$ & $\sim300$ & $1.2$ & $+29.7$ & $112$ & $1800$\\
\tableline
204949.78$+$000547.3\tablenotemark{b} & $19.66\pm0.03$ & $8660\pm144$  & $<0.169$ & $\lapprox\ 0.1$    & $1.7$ & $-26.0$ & $128$ & $1860$
\enddata
\tablecomments{$g$ (PSF) magnitudes are from SDSS DR6. $P_{\rm orb}$
is estimated by comparison to models from \cite{tauris99}; see also
\citet{vankerkwijk05}. Distances are calculated assuming a typical
$M_g = 8.5$. The listed DM is the maximum value used when searching
for pulsations; it corresponds approximately to twice the maximum value
obtained in the direction of each object with the \citet{cordes02} model
(for computing purposes the maximum DM is a multiple of $16$).  }
\tablenotetext{a}{Not included in \citet{eisenstein06} catalog.}
\tablenotetext{b}{\citet{kilic07a} identify J2049$+$0005 as an A2 star.}
\end{deluxetable*}

Figure~\ref{properties} gives log $g$ vs.\ T$_{\rm eff}$ for all of the candidate LMWDs as reported by \citet{eisenstein06}. We use the models in Figure~\ref{properties} to estimate M$_{\rm WD}$ for each of our LMWDs; their properties are presented in Table~\ref{props}. While the masses of these WDs are comparable to those of LMWDs detected as MSP companions, their temperatures are very different. For example, the companion to MSP J1012$+$5307, whose mass of $0.16\pm0.02$~M$_\odot$ makes it the lowest-mass confirmed WD in a MSP/WD system, has T$_{\rm eff} = 8550\pm25$~K \citep{vankerkwijk96}, which falls in the range for previously cataloged pulsar WD companions, $3500$~K~$<$~T$_{\rm eff} < 15,000$~K, regardless of WD type and mass \citep{hansen98, vankerkwijk05}. Indeed, none of the known He-core WD companions whose temperature has been measured is hotter than T$_{\rm eff} \approx 10,000$~K. Our LMWDs all have temperatures above $8000$~K, and most have T$_{\rm eff}\ \gapprox\ 15,000$~K, indicating that we are looking at a younger population of objects than that found to date as MSP companions\footnote{Because their colors are similar to those of main-sequence stars, WDs with T$_{\rm eff}\ \lapprox\ 5000$~K are unlikely to be targeted for/discovered via SDSS spectroscopy \citep{kilic06}.}.

A comparison to the evolutionary models produced by \citet{serenelli01} and \citet{althaus01} suggests that all but the lowest mass WD in our sample are younger than roughly $1$ Gyr, and that all are younger than $5$ Gyr (see Figure~\ref{properties}). These ages are calculated from WD cooling models, and correspond to the time since the end of mass transfer onto the WD companions. By contrast, typical MSPs have ages of $5$ to $10$ Gyr; these are estimated from the pulsar characteristic (spin-down) age and, in a few cases, also from the companion WD's cooling age \citep[e.g., J1012$+$5307;][]{lange01}.

\section{Green Bank Telescope Observations}\label{obs}
The LMWDs were observed with the GBT over the course of five nights
between 2006 January 12 and March 9. Each session began with a 1 min
observation of a test pulsar that was immediately analyzed to
confirm a correct observing setup; the integration times for each target
are listed in Table~\ref{props}. At the central observing frequency
of 820\,MHz, the Berkeley-Caltech Pulsar Machine analog/digital filter
bank \citep[the design is described in][]{backer97} provided 48\,MHz of
bandwidth split into 96 spectral channels, for each of which we recorded
total power samples every $72\,\mu$s. The resulting data were analyzed
using standard pulsar search techniques implemented in the PRESTO
package \citep{ransom01} installed in our computer cluster.
This analysis was similar to that described in \citet{camilo06}, with
appropriate ranges of trial dispersion measure (DM) to account for
dispersive interstellar propagation, and with substantial acceleration
correction to partially account for the changing pulsar period caused
by putative orbital motion.

We calculated the maximum DM expected in the direction of each target
using the \citet{cordes02} model for the distribution of free electrons
in the Galaxy.  A typical value for our high-Galactic-latitude targets
was $\approx\ 40$\,cm$^{-3}$\,pc.  In order to account for possible
uncertainties in the model, we de-dispersed the data up to a DM limit
twice that obtained from the model.  These values are given for each
target in Table~\ref{props}.

The orbital motions of the putative pulsars in these systems could
significantly affect the apparent pulsar spin period, particularly for the
companions to the lowest-mass WDs in our sample (known MSP/WD binaries
have orbital periods that range from a few hours to many months). In
Table~\ref{props}, we include rough estimates for binary periods based on
comparisons to the models of \cite{tauris99}. \cite{tauris99} calculated
orbital properties for $1$--$2$ M$_\odot$ companions to $1.3$ M$_\odot$
accreting NSs, and their models are consistent with the orbital periods
found for systems with less massive companions \citep{vankerkwijk05}. While for
$90\%$ of known pulsars the maximum orbital acceleration is $\leq |25|$
m s$^{-2}$ \citep{manchester05, joeri07}, for $40\%$ of our targets the 
estimated accelerations are $\gapprox\ 100$ m~s$^{-2}$. 
This significantly increases the parameter space
that needs to be searched for pulsed signals and hence the computing time
on the cluster, and also impacts our sensitivity limits (see discussion
in \S~\ref{acc_sens}).

\section{Discussion}\label{discussion}
No convincing pulsar signal was detected in our data. Below we discuss
the limitations of our search.

\subsection{Luminosity sensitivity}\label{lum_sens}

Models for WDs in the mass range of interest here predict absolute
magnitudes of M$_g \sim 8.5$ \citep{eisenstein06}. We use SDSS Data Release 6 photometry \citep[DR6;][]{DR6paper} to obtain rough
distance estimates to our LMWDs (see Table~\ref{props}).

Luminosity-wise, our searches were extremely constraining.  In order to calculate the minimum detectable period-averaged flux
density, we use the standard modifications to the radiometer equation.
We consider a pulsar duty cycle of 20\% (fairly typical of MSPs) and
compute sensitivity, conservatively, for $\mbox{DM}=50\,$cm$^{-3}$\,pc.
At 820\,MHz, the GBT gain is 2\,K\,Jy$^{-1}$ and the system temperature
is 25\,K.  The sky temperature varies with location, but at the relevant
high latitudes at this frequency it only adds a few Kelvin to the overall
temperature.  We consider an effective threshold signal-to-noise ratio
of 10.  For an integration time of 2400\,s (see Table~\ref{props}),
the sensitivity limit for long period pulsars is 0.16\,mJy.  Pulsar
luminosities are often measured at 1400\,MHz; using a typical spectral
index of $-1.7$, the limiting sensitivity at that frequency is about
$S_{1400} = 0.06$\,mJy.  For an MSP period of 3\,ms, our
sensitivity at 1400\,MHz was roughly 50\% worse, or 0.1\,mJy (and it
quickly degrades for shorter periods, being three times worse for 1\,ms).
(Twelve of our targets fall within the FIRST footprint; none 
are detected in these data, for which the sensitivity limit is roughly 
$1$ mJy \citep{first}.)

About 95\% of MSPs known in the Galactic disk have luminosities $L_{1400}
\equiv S_{1400} d^2 > 0.1$\,mJy\,kpc$^2$, and some $90\%$ have $L_{1400} >
0.5$\,mJy\,kpc$^2$.  Since our targets have typical estimated distances of
$1-2$ kpc, our $L_{1400}$ limits for 3\,ms periods ranged over $\approx\
0.1-0.4$ mJy\,kpc$^2$.  As far as luminosity alone is concerned,
therefore, we should have been able to detect approximately $90\%$ of known
MSPs had they been present in our targets and beaming radio waves toward
the Earth.

\subsection{Acceleration sensitivity}\label{acc_sens}
As mentioned earlier, if the lowest-mass WDs in our sample are in binaries
with NSs, these should be compact systems with orbital periods on the
order of a few hours to a few days \citep{tauris99, vankerkwijk05}. As
a result, pulse smearing due to orbital motions could occur, severely
complicating our search. To gauge the extent of this effect, we have
calculated orbital accelerations for a number of systems with LMWDs in
orbit around a $1.4$~M$_\odot$ NS. 

When conducting an accelerated pulsar search with PRESTO, the free
parameter related to the maximum acceleration searched actually specifies
the maximum drift of Fourier components that is corrected.
Relating this to a physical acceleration depends on the integration
time and on the putative pulsar period.  We consider a period of 3\,ms.
For our integration times and computing resources, it was impracticable
to fully correct for putative acceleration such as might be produced
in systems with orbital periods of about 8 hours or less.  For example,
for J0849$+$0445, a M~$\sim0.169$~M$_\odot$ WD that we observed for $3000$
s, the search sensitivity to a $3$ ms pulsar was limited to a maximum
acceleration of $60$~m~s$^{-2}$. Because this search alone required
several days of computing time on our $16$-node cluster, it simply was
not feasible to extend it to the maximum accelerations predicted for a
WD of its mass (e.g., $300$~m~s$^{-2}$ if $P_{\rm orb} \sim 3$ h and $1200$~m~s$^{-2}$ if $P_{\rm orb} \sim 1$ h).

Furthermore, for the six WDs with M~$\lapprox\ 0.169$~M$_\odot$, our
integration times may represent a substantial fraction of the binary
orbital period (e.g., as much as $50\%$ in the case of SDSS J1234; 
see Table~\ref{props}). In these cases, the assumption of a constant
apparent acceleration used in our search breaks down, further limiting
our ability to detect a pulsed signal from these targets.

Nevertheless, we were still very sensitive to most kinds of {\em known\/}
MSP/WD systems: essentially none are known in the disk of the Galaxy with
$P_{\rm orb} < 14$\,h.  In summary, while we were not very sensitive to
new kinds of exotic MSPs (e.g., with $P<1.5$\,ms, or with $P_{\rm orb} <
12$\,h), we should have detected some ordinary MSPs if they were orbiting
a substantial fraction of our targets.

\subsection{Constraining the number of LMWD/MSP systems}
\citet{joeri07} conducted a similar search for radio pulsations at $340$ MHz from unseen companions to eight SDSS LMWDs using the GBT. None of these observations contained signals warranting follow-up. 

The eight WDs targeted by \citet{joeri07}, described in \citet{liebert04}, were identified as candidate LMWDs based on initial model fits to SDSS spectra done by \citet{kleinman04}. However, of the eight candidates, three are no longer considered LMWDs based on more recent analysis of their spectra by \citet{eisenstein06}\footnote{These three WDs are SDSS J081136.34$+$461156.4, J102228.02$+$020035, and J130422.65$+$012214.2.}. We therefore updated the analysis performed by \citet{joeri07} in order to place more stringent constraints on the likelihood $P_{MSP}$ that a given LMWD has an MSP companion.

\citet{joeri07} assumed that a non-detection was the most likely outcome of their observations and estimated $P_{MSP}$ by using
\begin{equation}\label{eq_1}
(1 - P_{beam} \times P_{L} \times P_{acc} \times P_{eff} \times P_{MSP})^N > \frac{1}{2},
\end{equation}
where $P_{beam}$ is the MSP beaming fraction, $P_L$ is the luminosity completeness of the search, $P_{acc}$ the sensitivity to the systems' accelerations, $P_{eff}$ the search-algorithm success rate, and $N$ is the number of observed LMWDs. \citet{joeri07} find that $P_{MSP} <18\pm5\%$; this limit rises to $26\%$ if one removes the three objects no longer classified as LMWDs.

We use several of the same assumptions as \citet{joeri07} in estimating $P_{MSP}$. We set $P_{beam} = 0.7\pm0.2$ and chose $P_{eff} = 0.8$, a conservative estimate of our MSP-detection algorithm's success rate. Our luminosity sensitivity is $P_L = 0.9$ and our acceleration sensitivity is $P_{acc} = 0.9$; these two values reflect our confidence that our observations were sensitive enough to detect most MSPs with luminosities, orbital characteristics, and rotation periods comparable to those of the known population. We then find that $P_{MSP} < 10^{+4}_{-2}\%$.

\section{Conclusion}\label{concl}
We conducted a search for pulsar companions to $15$ low-mass white dwarfs at $820$ MHz with the GBT. No convincing pulsar signal was detected in our data. This is consistent with the findings of \citet{joeri07}, who conducted a search for radio pulsations at $340$ MHz from unseen companions to eight SDSS LMWDs (three of which are no longer considered LMWDs). However, our non-detections place stronger constraints on the probability that the companion to a given LMWD is a radio millisecond pulsar. Our observations allow us to lower this likelihood from $< 26\%$ to $< 10^{+4}_{-2}\%$. 

Any search for pulsars is inherently biased, and even a completely unbiased search does not preclude the existence of MSPs in these systems.  MSP beaming fractions are thought to be about $70\pm20\%$, and therefore in some sensitive searches MSPs have not been detected even though there is strong evidence for their presence \citep[e.g., the companion to the young pulsar J1906$+$0746;][]{lorimer06}. Given that NS blackbody emission is gravitationally bent, allowing us to view $>75\%$ of the NS surfaces in X rays \citep{belo02}, X-ray observations could yet detect MSPs that might exist in these systems. None of these LMWDs have been observed since the {\it ROSAT} All-Sky Survey \citep{voges99,fsc}, and none were detected in that survey.

If LMWD companions are not NSs, the likelihood is that their currently
unseen companions are also (very faint) WDs. Two dozen WD/WD binaries are
known; in $10$ systems both WD masses have been measured \citep[cf.][and
references therein]{nelemans05}. WD0957$-$666 has M$_1 = 0.37$
and M$_2 = 0.32$ M$_\odot$; WD1101$+$364 has M$_1 = 0.29$ and M$_2 =
0.35$ M$_\odot$. The other WDs in these systems have M $\geq 0.35$
M$_\odot$. In the $14$ other binaries discussed by \citet{nelemans05},
only one of the WD masses is measured, with the companion WD mass given
as a lower limit. Four of these systems harbor a WD with $0.23
\leq$ M $< 0.35$~M$_\odot$. If the SDSS LMWDs are confirmed
as WD/WD systems, they will represent a significant addition to this
population --- and in a mass range that is still poorly sampled.

\section*{Acknowledgments}
We thank Scott Ransom and the observing specialists at the GBT for their assistance and  Kurtis Williams and Craig Heinke for helpful discussions. M.\ Ag\"ueros is supported by an NSF Astronomy and Astrophysics Postdoctoral Fellowship under award AST-0602099.

The Robert C. Byrd Green Bank Telescope is operated by the National Radio Astronomy Observatory, which is a facility of the U.S.\ National Science Foundation operated under cooperative agreement by Associated Universities, Inc. We gratefully acknowledge GBT grant GSSP06-0008 for support of portions of our program.

Funding for the SDSS and SDSS-II has been provided by the Alfred P. Sloan Foundation, the Participating Institutions, the National Science Foundation, the U.S. Department of Energy, the National Aeronautics and Space Administration, the Japanese Monbukagakusho, the Max Planck Society, and the Higher Education Funding Council for England. The SDSS Web Site is http://www.sdss.org/.

The SDSS is managed by the Astrophysical Research Consortium for the Participating Institutions. The Participating Institutions are the American Museum of Natural History, Astrophysical Institute Potsdam, University of Basel, University of Cambridge, Case Western Reserve University, University of Chicago, Drexel University, Fermilab, the Institute for Advanced Study, the Japan Participation Group, Johns Hopkins University, the Joint Institute for Nuclear Astrophysics, the Kavli Institute for Particle Astrophysics and Cosmology, the Korean Scientist Group, the Chinese Academy of Sciences (LAMOST), Los Alamos National Laboratory, the Max-Planck-Institute for Astronomy (MPIA), the Max-Planck-Institute for Astrophysics (MPA), New Mexico State University, Ohio State University, University of Pittsburgh, University of Portsmouth, Princeton University, the United States Naval Observatory, and the University of Washington.


\begin{thebibliography}{37}
\expandafter\ifx\csname natexlab\endcsname\relax\def\natexlab#1{#1}\fi

\bibitem[{{Adelman-McCarthy} {et~al.}(2006)}]{DR4paper}
{Adelman-McCarthy}, J.~K., {et~al.} 2006, \apjs, 162, 38

\bibitem[{{Adelman-McCarthy} {et~al.}(2008)}]{DR6paper}
---. 2008, \apjs, 175, 297

\bibitem[{{Althaus} {et~al.}(2001)}]{althaus01}
{Althaus}, L.~G., {et~al.} 2001, \mnras, 323, 471

\bibitem[{{Backer} {et~al.}(1997)}]{backer97}
{Backer}, D.~C., {et~al.} 1997, \pasp, 109, 61

\bibitem[{{Becker} {et~al.}(1995)}]{first}
{Becker}, R.~H., {et~al.} 1995, \apj, 450, 559

\bibitem[{{Beloborodov}(2002)}]{belo02}
{Beloborodov}, A.~M. 2002, \apjl, 566, L85

\bibitem[{{Benvenuto} \& {De Vito}(2005)}]{benvenuto05}
{Benvenuto}, O.~G., \& {De Vito}, M.~A. 2005, \mnras, 362, 891

\bibitem[{{Camilo} {et~al.}(2006)}]{camilo06}
{Camilo}, F., {et~al.} 2006, \apj, 637, 456

\bibitem[{{Cordes} \& {Lazio}(2002)}]{cordes02}
{Cordes}, J.~M., \& {Lazio}, T.~J.~W. 2002, ArXiv e-prints, astro-ph/0207156

\bibitem[{{Driebe} {et~al.}(1998){Driebe}, {Schoenberner}, {Bloecker}, \&
  {Herwig}}]{driebe98}
{Driebe}, T., {Schoenberner}, D., {Bloecker}, T., \& {Herwig}, F. 1998, \aap,
  339, 123

\bibitem[{Eisenstein {et~al.}(2006)}]{eisenstein06}
Eisenstein, D.~J., {et~al.} 2006, \apjs, 167, 40

\bibitem[{{Hansen} \& {Phinney}(1998)}]{hansen98}
{Hansen}, B.~M.~S., \& {Phinney}, E.~S. 1998, \mnras, 294, 569

\bibitem[{{Hansen} {et~al.}(2007)}]{hansen07}
{Hansen}, B.~M.~S., {et~al.} 2007, \apj, 671, 380

\bibitem[{{Kilic} {et~al.}(2006){Kilic}, {Munn}, {Harris}, {Liebert}, {von
  Hippel}, {Williams}, {Metcalfe}, {Winget}, \& {Levine}}]{kilic06}
{Kilic}, M. {et~al.} 2006, \aj, 131, 582

\bibitem[{{Kilic} {et~al.}(2007{\natexlab{a}})}]{kilic07c}
{Kilic}, M., {et~al.} 2007{\natexlab{a}}, \apj, 671, 761

\bibitem[{{Kilic} {et~al.}(2007{\natexlab{b}})}]{kilic07a}
---. 2007{\natexlab{b}}, \apj, 660, 1451

\bibitem[{{Kleinman} {et~al.}(2004)}]{kleinman04}
{Kleinman}, S.~J., {et~al.} 2004, \apj, 607, 426

\bibitem[{{Koester} {et~al.}(1979){Koester}, {Schulz}, \&
  {Weidemann}}]{koester79}
{Koester}, D., {Schulz}, H., \& {Weidemann}, V. 1979, \aap, 76, 262

\bibitem[{{Lange} {et~al.}(2001){Lange}, {Camilo}, {Wex}, {Kramer}, {Backer},
  {Lyne}, \& {Doroshenko}}]{lange01}
{Lange}, C., {Camilo}, F., {Wex}, N., {Kramer}, M., {Backer}, D.~C., {Lyne},
  A.~G., \& {Doroshenko}, O. 2001, \mnras, 326, 274

\bibitem[{{Liebert} {et~al.}(2004)}]{liebert04}
{Liebert}, J., {et~al.} 2004, \apjl, 606, L147

\bibitem[{{Liebert} {et~al.}(2005)}]{liebert05}
---. 2005, \apjs, 156, 47

\bibitem[{{Lorimer}(2008)}]{lorimer08}
{Lorimer}, D.~R. 2008, Living Reviews in Relativity, 11, 8

\bibitem[{{Lorimer} {et~al.}(2006)}]{lorimer06}
{Lorimer}, D.~R., {et~al.} 2006, \apj, 640, 428

\bibitem[{{Manchester} {et~al.}(2005){Manchester}, {Hobbs}, {Teoh}, \&
  {Hobbs}}]{manchester05}
{Manchester}, R.~N., {Hobbs}, G.~B., {Teoh}, A., \& {Hobbs}, M. 2005, \aj, 129,
  1993

\bibitem[{{Marsh} {et~al.}(1995)}]{marsh95}
{Marsh}, T.~R., {et~al.} 1995, \mnras, 275, 828

\bibitem[{{Nelemans} {et~al.}(2005){Nelemans}, {Napiwotzki}, {Karl}, {Marsh},
  {Voss}, {Roelofs}, {Izzard}, {Montgomery}, {Reerink}, {Christlieb}, \&
  {Reimers}}]{nelemans05}
{Nelemans}, G. {et~al.} 2005, \aap, 440, 1087

\bibitem[{{Panei} {et~al.}(2007){Panei}, {Althaus}, {Chen}, \& {Han}}]{panei07}
{Panei}, J.~A., {Althaus}, L.~G., {Chen}, X., \& {Han}, Z. 2007, \mnras, 382,
  779

\bibitem[{{Ransom}(2001)}]{ransom01}
{Ransom}, S.~M. 2001, PhD thesis, Harvard University

\bibitem[{{Serenelli} {et~al.}(2001){Serenelli}, {Althaus}, {Rohrmann}, \&
  {Benvenuto}}]{serenelli01}
{Serenelli}, A.~M., {Althaus}, L.~G., {Rohrmann}, R.~D., \& {Benvenuto}, O.~G.
  2001, \mnras, 325, 607

\bibitem[{{Tauris} \& {Savonije}(1999)}]{tauris99}
{Tauris}, T.~M., \& {Savonije}, G.~J. 1999, \aap, 350, 928

\bibitem[{{Tauris} \& {van den Heuvel}(2006)}]{tauris06}
{Tauris}, T.~M., \& {van den Heuvel}, E.~P.~J. 2006, {Formation and evolution
  of compact stellar X-ray sources} (Compact stellar X-ray sources), 623--665

\bibitem[{{van Kerkwijk} {et~al.}(1996)}]{vankerkwijk96}
{van Kerkwijk}, M.~H., {et~al.} 1996, \apjl, 467, L89+

\bibitem[{{van Kerkwijk} {et~al.}(2005)}]{vankerkwijk05}
{van Kerkwijk}, M.~H., {et~al.} 2005, in Astronomical Society of the Pacific
  Conference Series, Vol. 328, Binary Radio Pulsars, ed. F.~A. {Rasio} \& I.~H.
  {Stairs}, 357

\bibitem[{{van Leeuwen} {et~al.}(2007)}]{joeri07}
{van Leeuwen}, J., {et~al.} 2007, \mnras, 374, 1437

\bibitem[{{Voges} {et~al.}(1999)}]{voges99}
{Voges}, W., {et~al.} 1999, \aap, 349, 389

\bibitem[{{Voges} {et~al.}(2000)}]{fsc}
---. 2000, VizieR Online Data Catalog, 9029, 0

\bibitem[{{York} {et~al.}(2000)}]{york00}
{York}, D.~G., {et~al.} 2000, \aj, 120, 1579

\end{thebibliography}
\end{document}